\documentclass[12 pt]{article}
\usepackage{amsmath, enumitem, setspace,graphics, comment,bm,bbm}
\usepackage[margin=1in]{geometry}

\usepackage{caption}
\usepackage[linesnumbered, ruled]{algorithm2e}
\SetKwRepeat{Do}{do}{while}%
\usepackage{amsfonts}
\usepackage{graphicx}
\usepackage{rotating}
\usepackage{amsthm}
\usepackage[longnamesfirst]{natbib}

\usepackage{hyperref}
\hypersetup{
    colorlinks=true,
    linkcolor=blue,
    filecolor=magenta,      
    urlcolor=cyan,
    citecolor=blue
}
\theoremstyle{definition}

\usepackage[svgnames]{xcolor}
\usepackage{tikz}
\usepackage{bm}
\usetikzlibrary{fit,positioning}
\usepackage{rotating}
\usepackage{csquotes}
\usepackage{subcaption}
\usepackage{listings}

\title{Machine Learning for Public Administration Research, with Application to Organizational Reputation}

\author{L. Jason Anastasopoulos\\ 
				\texttt{ljanastas@uga.edu} 
				\and
			Andrew B. Whitford \\
			 \texttt{aw@uga.edu}
			 }

\begin{document}
\maketitle

\doublespacing
													\singlespacing
													
\begin{abstract}
\noindent  
Machine learning methods have gained a great deal of popularity in recent years among public administration scholars and practitioners. These techniques open the door to the analysis of text, image and other types of data that allow us to test foundational theories of public administration and to develop new theories. Despite the excitement surrounding machine learning methods, clarity regarding their proper use and potential pitfalls is lacking. This paper attempts to fill this gap in the literature through providing a machine learning ``guide to practice'' for public administration scholars and practitioners. Here, we take a foundational view of machine learning and describe how these methods can enrich public administration research and practice through their ability develop new measures, tap into new sources of data and conduct statistical inference and causal inference in a principled manner. We then turn our attention to the pitfalls of using these methods such as unvalidated measures and lack of interpretability. Finally, we demonstrate how machine learning techniques can help us learn about organizational reputation in federal agencies through an illustrated example using tweets from 13 executive federal agencies.

\end{abstract}


\cleardoublepage

\doublespacing
\pagenumbering{arabic}

\section{Introduction}

Perhaps the most important problem of the next several decades will be how humans cope with unprecedented increases in the volume of data. A 2012 report by IDC for the EMC Corporation projected that the ``digital universe'' will reach 40 zettabytes (ZB) by 2020 ~\citep{gantz2012digital}. A zettabyte is one sextillion bytes; 40 ZB is 40 trillion gigabytes. Obviously, that was over half a decade ago so it is likely we have exceeded that forecast by now, but the importance of the prediction is not the level but its trajectory. That forecast itself exceeded previous forecasts by 14 percent; the predicted increase in the size of the digital universe is fifty-fold from 2010 to 2020. In 2006 the entire world's hard drives held 160 exabytes ~\citep{reinsel2007expanding}. The increase alone from 2011 to 2012 was estimated to be 48 percent (to 2.7 ZB)~\citep{predicts2012will}. 

Yet, ``in an information-rich world, the wealth of information means a dearth of something else: a scarcity of whatever it is that information consumes''; information consumes attention ~\citep[40]{simon1971designing}. For Simon, thinking about attention was like thinking about information itself. The bit is a way of thinking about units of information; there are eight bits to the byte. But there was no easy, parallel way of measuring attention. This was because humans are serial processors - a point that Simon made by comparing humans to the machines of that era, machines that were also serial processors. Simon's concern was how to overcome this natural and inevitable attention deficit given the likelihood that data would explode. 

Data \emph{qua} data is a curious focus in most cases. The real question is how we collect and process data - how data are transitioned first into information and then, perhaps, knowledge. For Herbert Simon, computers as ``thinking machines'' were part of this answer but only if designed to improve the situation. In the 1971 speech quoted here, Simon's principle was simple: a machine (just like any designed entity such as an organization) would help with the attention problem only if ``it absorbs more information previously received by others than it produces - that is, if it listens and thinks more than it speaks''~\citep[42]{simon1971designing}. In a response to this principle that Simon offered in his speech, the organizational economist Martin Shubik told the story of the owl who advised a centipede, who complained of sore feet, to walk an inch off the ground for the next two weeks. The centipede saw the value of the advice but wanted to know how to achieve it; the owl replied ``I have solved your conceptual problem. Don't bother me with the technical details''~\citep[61]{simon1971designing}.  

This tension between concept and implementation has been present for several years in the proclamation that big data will transform the practice and theory of governance. As with the bit inside the byte inside the zettabyte, an important but overlooked aspect of the big data revolution is that a datum - big or small - has little to no value in and of itself. Indeed, the value of the information offered by big data does not necessarily relate to the size of the data but rather from our ability to use and interpret big data in ways that are meaningful for public administration researchers and practitioners. Simon's problem of finding an optimal processing system (either a machine or an organization) is even harder when the world is awash with data.

The purpose of this paper is to establish machine learning (ML) algorithms, a group of data-driven empirical methods that combine insights from computer science and statistics, as the most likely next generation of mechanisms for unlocking the potential of big data. These algorithms provide methods for handling the variety and volume of data constantly being generated by public organizations~\citep{qiu2016survey}. Moreover, they are already in use by public managers across a broad array of organizations who use them to make everyday decisions~\citep{helsby2018early}. 

We review these techniques because they offer respite from the flood of data in a world of limited (and perhaps shrinking) attention. Yet, despite the promise that these techniques hold for harnessing the power of big data, we believe that misunderstandings about their proper usage and limitations can potentially lead to the deterioration of public organizations and a decline in research quality over time. There are benefits from developing a basic understanding of these algorithms at the research level and also at the implementation level. This understanding includes knowledge about what the algorithms can do and cannot do; it includes, as best possible, an understanding of how the algorithms work (although, to be honest, in some cases we really do not know how they work). The benefits will only increase as reliance on these methods will inevitably spread dramatically in the very near future.

This paper provides an introduction to ML algorithms for public administration research and practice, highlighting both the promise that these methods hold and the perils that they pose through improper use.  The next section provides a broad overview of ML algorithms and a set of principles to guide their usage. The following section reviews an application and shows how machine learning can be used to study theories of reputation in executive agencies. Section 4 discusses three perils of improperly using these algorithms and three promises they hold for the future of public administration research and practice. The last section returns to the theme of data availability and velocity given constraints on human attention, and discusses the roles that machine learning and artificial intelligence will have in the future of research and of governance in public organizations.

\section{ML as a Solution to an Emerging Problem}

In a nutshell, ML algorithms are a group of techniques that are able to effectively learn from data. Although these techniques are grounded in statistical theory,  they are often referred to as algorithms rather than models because their ability to ``learn'' from data always involves a computational component. At a more figurative level, ML can be thought of as the art and the science of combining statistics and computer science to produce accurate predictions from a wide variety of data sources. These predictions, in turn, can help handle the attention problem in data-rich environments that Simon focused on - by scholars who want to gain a richer understanding of bureaucratic and organizational behavior, and by public administrators who are searching for ways to enhance decision-making processes. 

\subsection{ML Basics}

The goal of statistical learning, the theory which underlies ML algorithms, is the estimation of a function of the data $X$ that will make optimal predictions about some outcome $Y$ as shown in Equation~\ref{eq:1}:

\begin{equation}\label{eq:1}
			Y = f(X) + \epsilon
\end{equation}
		        
\noindent Specifically, all of statistical learning theory can be boiled down to (a) posing a problem and then (b) answering a question that follows from that problem. The problem centers on finding a set of predictions $\hat{Y}$ and an estimated function $\hat{f}(X)$, as is shown in Equation~\ref{eq:2}: 

\begin{equation}\label{eq:2}
			\hat{Y} = \hat{f}(X)
\end{equation}
					        
\noindent The question that follows from this involves the problem of ``best.'' Consider how in the case of ordinary least squares (OLS) we are comforted by proofs of ``BLUE'' (that under specific conditions OLS is the best linear unbiased estimator) and in the case of maximum likelihood estimation that we know that the algorithm converges to the answers we seek due to asymptotics. 

If we step back and take a more general viewpoint, though, we can pose this following question: what is the ``best''  function of the data $f(X)$ for generating the most accurate predictions $\hat{Y}$ (where ``best'' is defined as the function which minimizes the average difference between the true outcomes $Y$ and the predicted outcomes $\hat{Y}$), as shown in Equation~\ref{eq:3} below:

\begin{equation}\label{eq:3}
			\arg\min_{\hat{f}(X)} \mathbb{E}[(Y - \hat{Y})] = \arg\min_{\hat{f}(X)} \mathbb{E}[Y - \hat{f}(X)]
\end{equation}

\noindent Focus on the data $Y$ and $X$ and the predicted outcomes $\hat{Y}$. The outcome $Y$ and the predicted outcomes $\hat{Y}$ would ordinarily be referred to as the dependent variable and the predicted values of the dependent variable, respectively. In an ML context, however, the dependent variable is often referred to as the \textit{target} when the variable is either continuous or categorical or the \textit{class label} when the variable is categorical. Similarly, in more familiar statistical contexts, the data predicting the target $X$ (either a matrix or a vector) is ordinarily referred to as the \textit{covariates}, \textit{predictors}, or \textit{independent variables}. In an ML context, however, these data are always referred to as the \textit{features} of the data. 

\subsection{Supervised and Unsupervised Learning}

Putting all of this together, ML algorithms can be broadly conceptualized as two general types: supervised and unsupervised~\citep{james2013introduction}. The distinction between these two types is defined by the nature of the target (the $Y$ variable). Table~\ref{tab:supunsup} provides a brief description of each type of algorithm along with some relevant examples.

\begin{table}[ht!] \footnotesize
\centering
\begin{tabular}{llll}

\hline \hline
			\textbf{Algorithm} & \textbf{Description} & \textbf{Human Interpretation}  & \textbf{Examples}\\ \hline
			Unsupervised & Generates outcomes/types & Required to determine   & Topic models \\
									& $Y$  directly from predictors/ &  what the types discovered by & K-Means Clustering \\
									& independent variables $X$ &		the algorithm mean            & Principal-Components Analysis/ \\  
																			& & & Factor Analysis \\		
																			& & & \\		
			Supervised & Predicts outcomes/types on 		& Often required during	& Decision tree methods \\
			 					& new data by estimating a model & the coding of the $Y$ variable. & Naive Bayes \\
			 					& $Y=f(X) + \epsilon$ and applying& Required to assess algorithm & Neural networks \\
			 					& the model to new data to generate & performance 					& Support vector machines \\ 
			 					& predictions $\hat{Y} = \hat{f}(X)$	 &  &									Lasso \\ \hline \hline		
																	
\end{tabular}
\caption{Unsupervised and supervised machine learning algorithms}
\label{tab:supunsup}
\end{table}

\subsubsection{Unsupervised Techniques}

Broadly speaking, unsupervised machine learning techniques are useful primarily for uncovering latent patterns or groupings within data when there is no measured outcome. In public administration research, these kinds of techniques are best applied to circumstances in which there is a clear a priori connection between a theoretical concept or quantity being measured in the data, but it is not clear how that concept will manifest. 
Examples of applications of these techniques include the use of an unsupervised method known as ``topic models'' to uncovering latent theoretical concepts from documents produced by governmental organizations. 
For example ~\cite{hollibaugh2018use} uses topic models to measure changes in the priorities of federal agencies using documents generated by these agencies.  Similarly,~\cite{anastasopoulos2017computational} use county budget statements from California to measure and test~\cite{schick1966road} 's theory of budgeting functions and priorities. 
In political science research~\cite{roberts2014structural} use topic models to extract politically relevant information from open ended survey responses. 

\subsubsection{Supervised Techniques}

For public administration research, the goal of supervised learning techniques is to replicate human classification or measurement of some outcome where the general rules regarding how the outcomes were measured are unclear or incomplete in some sense. A canonical example of this is sentiment analysis, such as identifying whether documents contain positive or negative sentiment toward a product like a film. A human reader can easily identify a negative or positive movie review. However, the semantic complexities of texts often contain elements of sarcasm, so transforming the inherent human ability to identify negative and positive reviews into a series of lingustic rules exceedingly difficult and results in  poor performance when these rules are applied to new texts. It is hard to classify as ``negative'' reviews containing negatively-valanced  words (e.g., ``bad'', ``awful'') and as ``positive'' reviews containing positively-valanced words (e.g., ``good'', ``great'').

On the other hand, supervised learning algorithms, can use automated methods to derive linguistic rules to distinguish between these categories with a much higher probability of success than rules arrived at through human based deduction, despite the fact that humans can inductively classify these texts with ease through reading the documents directly.  

From the perspective of public administration research, supervised methods should be used in two situations: (1) where human-coded data (a set of features/covariates and some outcome/target) are already available and the researcher seeks to measure or predict that outcome on a much larger data set, and/or (2) when human coded labels/measures can be generated with relative ease conditional on more detailed theoretical guidance.

Examples of the first case include ~\cite{bonica2016inferring}, which uses supervised machine learning to infer ideological scores (outcome) based on roll call voting in Congress using campaign contribution data (features), and ~\cite{anastasopoulos2018understanding}, which predicts delegation and constraint ratios (outcomes) using the texts of European Union legislation (features).\footnote{They employ a dataset created by~\cite{franchino2007powers} that contains hand-coded legislation, and use machine learning to replicate the (implicit) rules used to hand-code the legislation and apply these machine learning-generated rules to a much larger database of legislation across time.} As an example of the second case is ~\cite{anastasopoulos2018job}, which uses machine learning and relatively simple coding instructions provided to human coders to identify low and high-skilled classified ads from the Miami Herald before and after increases in Cuban immigration to Miami brought about by the Mariel Boatlift. 

In this paper, our classification of tweets described below are conducted along the lines of the second point. 
Here, we have clear theoretical guidance provided to us by prior work done by~\cite{carpenter2012reputation} and apply this theoretical guidance to tweet classification for the purpose of measuring and understanding organizational reputation.

Supervised learning algorithms can also be used for a variety of other applications in the public administration context where theory is well developed or there is pre--existing data that can be leveraged. For example we might be interested in predicting the performance of public managers (target) using measures of internal management (features)~\citep{favero2014goals}. This would be a case of supervised learning because the model that we have estimated or \textit{trained} for the purpose of predicting the performance of public managers uses both the performance data itself and the internal management metrics to build a prediction machine that can learn about performance metrics from internal management measures. In contrast, unsupervised learning algorithms infer information about the target using only the features. For instance, given only measures of internal management as defined by~\cite{favero2014goals}, can we construct a metric that accounts for the differential performance of public managers along some latent dimension? In the case of performance among public managers, this dimension might be be something like ``competence'' or ``experience.''

\subsubsection{Machine Learning and ``Big Data''}

In a more general sense, outside of the areas of text analysis, ML can be incredibly useful when dealing with ``big data'' problems in two ways: (1) dimensionality reduction and; (2) principled model selection.
One of the most significant challenges faced by researchers using big data includes either possessing a large number of variables, a large number of observations or both. 
This is especially problematic in situations in which it is unclear how the theoretical underpinnings of a model map onto measured variables. ML is particularly useful in these situations  because it allows the researcher to reduce the dimensionality of their data (ie select variables) with automated methods. 
For example, recent research on public service motivation (PSM) by~\cite{van2016public} has shed light  on the personality traits linked to PSM. For researchers interested in PSM in different contexts, unsupervised ML methods such as k-means clustering or factor analysis, can be used to extract personality clusters from surveys of public sector workers in which a large range of questions about personal beliefs and values are asked.
Related to dimensionality reduction is the question of ethical model selection. When choosing variables to include or exclude in an ordinary regression model when little theoretical guidance exists, the temptation to ``p-hack,'' that is, base model selection on the extent to which statistical significance improves on relevant covariates as a function of the inclusion of covariates, increases dramatically. ML methods such as the LASSO and the Bayesian LASSO provide automated, impartial model selection in these contexts and thus can potentially reduce the likelihood of p-hacking with big data~\citep{bloniarz2016lasso,ratkovic2017sparse}.

While both supervised and unsupervised machine algorithms have many applications and great potential in public administration research, we focus on supervised machine learning in this paper because we believe that supervised learning algorithms comprise the most useful set of tools -- and also that these are the most difficult to apply and to interpret. We continue this overview with a discussion of the basic elements of supervised learning (training, testing, cross-validation and performance measurement). We conclude with a brief discussion of an essential theoretical concern in this area: the bias-variance trade-off. 

\subsection{Training, Validation, and Testing}

The goal of supervised learning is to build a prediction machine that can handle our chosen task. For instance, in the application below, we describe in detail how supervised ML can help us learn about organizational reputation. Specifically, we want to know how agencies use social media to build or maintain each of the four types of organizational reputation defined by~\cite{carpenter2012reputation}, as shown in Table~\ref{tab:carpkrrep}.

\begin{table}[ht!] \footnotesize
\centering
\begin{tabular}{l}

\hline \hline	
Performative Reputation \\
Moral Reputation \\
Procedural Reputation \\
Technical Reputation \\
\hline \hline	

\end{tabular}
\caption{Types of Organizational Reputation}
\label{tab:carpkrrep}
\end{table}

At a very basic level, in our example we are interested in determining the extent to which tweets produced or retweeted by 13 federal executive agencies can be identified as containing content related to each of these categories. A supervised machine learning approach to this problem requires performing the following tasks:

\begin{enumerate}
\item \textbf{Hand coding} -  For a subset of tweets, code by hand which tweets (for a given agency) are related to each of the four categories (or no category at all). 

\item \textbf{Training} - Randomly parse the hand-coded data into two subsets.  The ``training'' subset will be used to ``teach'' the algorithm the words and phrases that relate to each category. The other subset is a ``testing'' or ``holdout'' subset.\footnote{If enough data are available, data will be partitioned into a third ``?validation'' subset  which is used to ``fine tune'' the parameters of the algorithm to ensure the best possible performance~\citep{james2013introduction}} 

\item \textbf{Testing} - After training the algorithm, the final performance of the algorithm is assessed on the testing subset -- fresh data on which it has been neither trained nor validated. In this case, performance is ultimately a measure of the classifier's ability to accurately classify a tweet according into one of the four categories.
\end{enumerate}
 
The payoff to all of this is that, after testing performance is assessed, the trained classifier can be used as a ``prediction machine.'' We can apply this machine to a much larger dataset than the one that was originally hand-coded. In the parlance of Simon, the solution to our attention problem is a thinking machine.

We demonstrate in our application that an ML algorithm known as gradient boosted trees, a type of decision tree method, can be trained to identify the moral reputation content referred to in ~\cite{carpenter2012reputation} in a database of 26,402 tweets. We train this ML algorithm using a collection of human-coded tweets that comprise less than $1\%$ of this database (only 200\footnote{Due to a  technical error on Mechanical Turk, one tweet from the original 200 sampled database was excluded from being coded and included in the final trained model.} tweets).\footnote{A natural question is why divide the coded data into these two partitions when you could theoretically use all of the hand coded data to train the machine learning algorithm? We discuss a very specific and important theoretical reason for this known as ``overfitting'' below in the section about the ``bias-variance'' tradeoff.}
	
Before turning to a description of the different types of measures that are used to assess the performance of machine learning algorithms more generally, it is useful to take a step back to consider exactly what a supervised machine learning algorithm is doing when it is being used to classify data. From the above steps, it is clear that the legitimacy of our entire project depends upon how well the training data are coded. If the training data are coded with care, then the classifier will be able to pick up on the linguistic distinctions in different labeled reputation categories. However, if the training data were coded haphazardly and with little care, ML classifiers will not be able to successfully detect distinctions between tweets of different categories. 

The takeaway here is that there is no substitute for high quality data. Machine learning algorithms can only perform well if the data that we feed to them is of high quality. As we discuss below, while the ability to learn from large pools of data is a key promise of this family of techniques, an accompanying peril is that any machine learning algorithm is only as good as the data on which it is trained. 

\subsection{Loss Functions and Performance Metrics}

In a nutshell, the training process for building a ML algorithm that generates predictions in an optimal manner involves partitioning data into training and test sets. But what does optimal mean, and how we can operationalize this definition? We want to recognize the seeming circularity of this enterprise. We want to both (a) maximize the performance of our ML algorithm and (b) choose an algorithm for the problem we are trying to solve that is the best algorithm for that particular problem. 

Clarifying these points requires a discussion of loss functions and performance metrics. Loss functions help us estimate the parameters of ML algorithms to get the best possible predictions. We use performance metrics to judge how well a particular ML algorithm performs a given prediction task.
	
The literature here defines two general types of prediction problems~\citep{james2013introduction}. The first type is regression problems, which involve the prediction of continuous variables. An example of a regression problem might be predicting a managerial employment score; another might be predicting a budget for a particular fiscal year. The second type is classification problems, which involve predictions about the class, or category, into which an observation falls. In our illustrative example below, we offer an algorithm that predicts the tweets that are related to the class defined as the moral reputation of an executive agency. The distinction between regression and classification problems is important when we decide (a) what criteria we will use to estimate the parameters of the model during the training process and (b) how we will measure the performance of the trained algorithm. 
	
When we make a regression prediction, we want to estimate the parameters so that the predicted values (as given by our model) and the actual values (on which the model is estimated) are as close as possible. Consider a prediction problem in which we estimate a model by OLS that can predict how much will be allocated to public services in an annual budget at the county level, where $B$ might be the dollar amount of the annual budget outlays for public services and $X$ would be a matrix of predictors (e.g., demographic, political, and other factors measured at the county level):

$$
B = \beta_0 + \beta_{1} X_{1} + \cdots \beta_{p} X_{p} + \epsilon
$$

To estimate the parameters to produce the best possible predictions, we need to select parameter values that minimize the difference between the actual values  $B$ and the predicted values. This difference can be expressed as the OLS mean squared prediction error (MSE):
	
\begin{equation}\label{eq:lossmse}
\displaystyle MSE  = \sum_{i=1}^{N} \frac{ (B_{i} - \hat{B}_{i})^{2}}{N}
\end{equation}

\noindent where $\hat{B}_{i}$ are  the predicted values from the estimated regression model:

$$
\hat{B}_{i} = b_0 + b_{1} X_{1} + \cdots + b_{p} X_{p} 
$$

\noindent Removing the denominator from Equation~\ref{eq:lossmse} gives our cost or \textit{loss function}, $J(\beta)$:

\begin{equation}\label{eq:jbeta}
J(\beta) = \sum_{i=1}^{N} (B_{i} - \hat{B}_{i})^{2}
\end{equation}

Equation~\ref{eq:jbeta} informs how we select the parameter values $\beta$. It is composed as a function of the parameter values $\beta$ that we seek to estimate. Because it is convex, we can estimate parameter values by taking derivatives of $J(\beta)$ with regard to each of the parameters, set each derivative equal to zero, and then solve those equations to find the parameter values:

\begin{equation}\label{eq:jbetamin}
\arg\min_{\beta} J(\beta) = \arg\min_{\beta}  \sum_{i=1}^{N} (B_{i} - \hat{B}_{i})^{2}
\end{equation}

\noindent Note that this function is built on squared differences. In the ML literature, the loss function built on squared differences is referred to as $L_2$ (squared loss). Yet, $L_2$ loss is just one loss function that defines how we treat the differences between the true and predicted values. In contrast, the $L_1$ loss function is built on the absolute values of the differences between the observed and predicted values:

\begin{equation}\label{eq:jbetaabs}
J(\beta) = \sum_{i=1}^{N} \left | B_{i} - \hat{B}_{i} \right |
\end{equation}

\noindent The point here is that loss functions are important mechanisms for assessing the performance of an ML algorithm. Whether we focus on $L_1$ or $L_2$ loss, though, is a harder issue that depends on the context of the machine learning problem because the decision can ultimately affect the performance of the algorithm~\citep{bousquet2008tradeoffs}.

Classification problems for categorical variables are slightly more complicated because we rely on predicted probabilities. Consider the budget example from described above, but instead frame it as a classification problem by coding as 1 any decrease in a budget from time $t-1$ to $t$; code increases or no changes as 0. A machine learning classifier would help us predict a budget decrease in the next fiscal year by building a logistic regression:

\begin{equation}\label{eq:losslogit}
		logit[\mathbb{E}(B | X)] = \beta_0 + \beta_{1} X_{1} + \cdots + \beta_{p} X_{p} + \epsilon
\end{equation}

\noindent and the predicted probabilities for the trained model would be:

$$
  		      \hat{p}_{i} =  p(B_{i} = 1 | X_{i}) = \frac{1}{1 + exp-(b_0 + b_{1} X_{1i} + \cdots b_{p} X_{pi})}
$$

\noindent These predicted probabilities could then classify new data as budget reductions if $\hat{p}_{i} > 0.5$ and increases or no changes if $\hat{p}_{i} < 0.5$.
 
Our loss function should account for the probabilities produced ($\hat{p}_{i}$), the predicted class label from those probabilities ($\hat{Y}$), and the true class label ($Y$). A popular loss function for this situation is the  \textit{cross-entropy loss function}:

\begin{equation}\label{eq:celoss}
		 \sum_{i=1}^{N}  - [y_{i} log(\hat{p}_{i})+(1- y_{i})log(1- \hat{p}_{i})]
\end{equation}

\noindent where $y_{i}$ is a binary indicator that equals $1$ if the class label is correctly identified and equals $0$ if the class label is not correctly identified; $\hat{p}_{i}$ is the predicted probability as defined above.

This describes the criteria used by ML algorithms to estimate models with the highest amount of predictive success. We still need a way to measure how well the algorithms perform on the test data so that we can choose the best algorithm for the problem at hand; we also need it to fine tune a chosen algorithm to ensure maximum prediction success. 

For regression problems, we can use the mean squared or root mean squared prediction error on the test data (MSE/RMSE test) to assess model performance. For classification problems, we can combine three metrics into a single measure called the F1 score to assess classifier performance; these three metrics are accuracy ($a$), specificity ($\sigma_1$), and sensitivity ($\sigma_2$). MSE/RMSE are familiar; measuring performance for classification problems is more complicated. 

Consider measuring classifier performance by the accuracy of the classifier's predictions: the proportion of times that the classifier is correct when it makes a prediction about a binary class label. For our binarized budget variable, perhaps we trained a classifier, assessed its final performance on test data, and discovered that the classifier predicted when a budget would decrease, remain the same or increase 95 percent of the time (it had a 95 percent accuracy rate). This sounds like an accurate classifier, but if only 5 percent of the test data had budgets that decreased, the classifier can achieve 95 percent accuracy rate by simply predicting that no budgets will decrease (the modal category). A useful classifier must be able to detect when a budget will decrease. So we turn to other performance metrics that tell us how well the classifier distinguishes between true and false positive results (precision) and how good a classifier is at detecting positive class labels (recall).

Precision is a measure of the true positive rate and provides us with the answer to the following question: of the observations that the classifier identified as containing a positive class label, what \% of them are correctly identified as such? \textit{Recall} or \textit{sensitivity} measures the \textit{true positive} rate of a classifier. It provides us with the answer to the question: of all of the observations that have a positive class label, what  proportion of them will be detected by the classifier? Finally, the $F_1$ score gives us a measure of overall ``quality'' of the classifier in terms of its ability to correctly identify and detect observations with positive class labels. Table~\ref{tab:acsesp} provides technical details.

\begin{table}[hbpt!]
\centering
    \begin{tabular}{lll}
\hline \hline
    	\textbf{Metric}  & \textbf{Formula} & \textbf{Description} \\ \hline
    	   &  & \\
    	Accuracy &  $\frac{tp + tn}{tp+tn+fp+fn}$ & Probability of correctly classifying observations. \\
    	   &  & \\
    	Precision   &  $  \frac{tp}{tp+fp}$ & Probability of correctly classifying positive observations.  \\
    	 &  & \\
    Recall  &  $ \frac{tp}{tp+fn}$ & Probability of detecting a positive observation. \\ 
    &  & \\
    $F_1$ & $2 \times \frac{Precision \times Recall}{Precision + Recall}$ & Overall measure of classifier quality. \\
    \hline \hline
    \end{tabular}
   \caption{Performance metrics for determining which classifier is best suited for a binary classification task.}
   \label{tab:acsesp}
\end{table}




\subsection{The Bias--Variance Tradeoff}

The bias--variance tradeoff, a foundational concept in statistical learning theory, provides a basis for assessing the performance of ML algorithms and a guideline for training them for classification and regression tasks~\citep{geman1992neural}. Recall that the most general way of thinking about the performance of any ML algorithm is by measuring the expected difference between the true values ($Y$) and the values estimated from the data ($\hat{Y} = \hat{f}(X)$):

$$
\mathbb{E}[Y - \hat{f}(X)]
$$

\noindent Also, recall that the goal of ML is to estimate a function $\hat{f}(X)$ that minimizes this error. This error can be decomposed into:

\begin{equation}\label{eq:varbi}
\mathbb{E}[Y - \hat{f}(X)] = Var[\hat{f}(X)]  + \left[Bias[\hat{f}(X)]\right]^{2} + Var(\epsilon)
\end{equation}

\noindent In Equation~\ref{eq:varbi}, $Var[\hat{f}(X)]$ is the variance of the predictions, $Bias[\hat{f}(X)]$ is the bias of the predictions, and $Var(\epsilon)$ is the variance of the error term. We cannot do much about $Var(\epsilon)$ since this term depends upon unobservables, so we focus on the variance of the predictions and the bias of the predictions. 

Consider the variance of the predictions ($Var[\hat{f}(X)]$). This measures how much $\hat{f}$ would change when applied to a different data set. More specifically in the case of ML, $Var[\hat{f}(X)]$ tells us how well an algorithm that was trained on one test dataset would perform on new datasets. Higher values suggest poorer performance; lower values suggest the opposite.

Now consider the bias of the predictions ($Bias[\hat{f}(X)]$). This measures how well the model fits the data at hand; in the ML context, this is the training data on which we trained our algorithm. High levels of bias suggest that our model is not sufficiently complex to fit the data -- or that the data are of poor quality. 

Table~\ref{tab:varbi} provides insight into tradeoff between the bias and the variance of a model -- one of the most significant issues faced in ML. Consider the OLS model: $Y  = \beta_{0} + \sum_{i=1}^{p}\beta_{i} X_{i}+ \epsilon$. For the most part, when $p=1$, this model will poorly predict $Y$ using the training data (high bias) and also poorly predict $Y$ using the test data (high variance). Increasing the number of independent variables and parameters ($p = 2, 3, 4, \ldots $) will reduce model bias (as the model better fits the training data), but eventually the model fits the training data so well that it is not useful for predicting data outside of the training data (called ``overfitting''). This is the essence of the bias--variance tradeoff: there is always a tradeoff at a certain point between how well a model makes predictions on the training data and how well the model makes predictions on the test data. 

\begin{table}[ht!]
\centering
  \begin{tabular}{lll}
  \hline\hline 
   			 $Var[\hat{f}(X)]$ & $Bias[\hat{f}(X)]$ & Implication \\ \hline
				\textit{High}  & \textit{Low} & Overfitting \\
				\textit{High}  & \textit{High} & Underfitting \\ 
				\textit{Low} & \textit{High} & Underfitting \\ 		
				\textit{Low} & \textit{Low} & Optimal \\ \hline\hline 
  \end{tabular}
  \caption{Implications of different levels of variance and bias}
  \label{tab:varbi}
\end{table}

\begin{figure}[ht!]
\centering
\includegraphics[width = 0.75\textwidth]{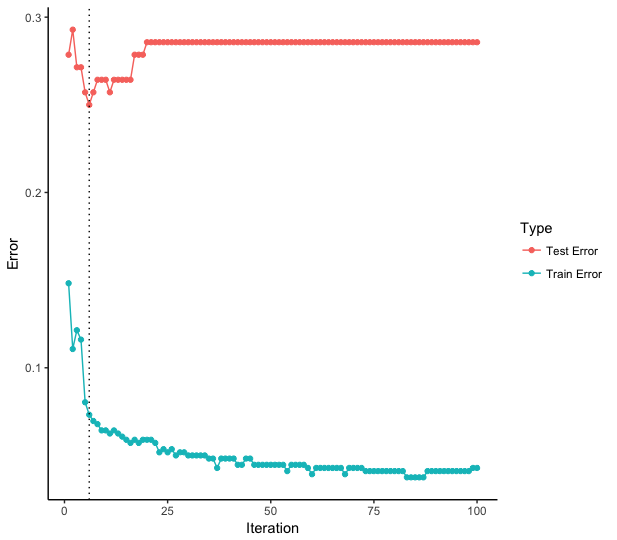}
\caption{An illustration of the bias-variance tradeoff}
\label{fig:biasvarfig}
\end{figure}

Figure~\ref{fig:biasvarfig} shows the bias--variance tradeoff present when training a gradient boosted trees classifier to identify moral reputation content in tweets. While the fit of the model on the training data continues to improve, the model's fit on the test data becomes \textit{worse} after the point marked off by the dotted line (iteration 6). While the test error reaches a minimum at the dotted line (iteration 6) and increases after that, the training error continues to decrease. \textit{An ideal classifier will have low variance and low bias, but there is always a tradeoff between the two.}

This section has provided an overview of the basics of ML, along with extended discussions of the differences between supervised and unsupervised learning, the nature of training and model validation, the importance of loss functions and performance metrics, and the bias--variance tradeoff. These are broad discussions of the overall ML enterprise, but they mask the very specific modeling choices that researchers incur in the context of a research project. To better focus attention on those choices and the differences between the ML approach and traditional techniques, we next turn to an extended application: the use of a learning algorithm to detect organizational reputation in federal agency Twitter feeds.

\section{Application: Learning about Reputation in Agencies}

Public administrators face constant external challenges to their viability as they balance accountability to politicians with their public service mission. In their 2011 \textit{Public Administration Review} article, ~\cite{carpenter2012reputation} identify three key challenges, including the maintenance of broad--based support for the agency and its activities, the handling of potential enemies and/or disaffected supporters of the agency, and projecting an image of ``consistency and flexibility.'' They argue that the ability of public administrators to deal with these challenges depends a great deal on ``organizational reputation.''

Public administrators can use different strategies for each of these challenges, but increasingly those strategies require the agency to engage with the public through  media. Historically, this included print or television news media, or communication via agency press releases and internal documents issued by the agencies themselves. Now the Internet and social media are fundamentally transforming the relationships between agencies and the individuals and groups they serve -- as well as changing how agencies communicate when addressing the three challenges identified above. 

Social media allows agencies to  communicate frequently to many different actors as a way of maintaining support and demonstrating relevance. Agencies can use social media to address disaffected supporters (or potential enemies) by communicating directly with them on those platforms. Finally, agencies can  project images of ``consistency and flexibility'' through content they produce daily on platforms such as Twitter. 

Our conjecture is that social media can play an important role in understanding organizational reputation. To assess this conjecture, we demonstrate how to use a supervised machine learning method known as ``gradient boosted trees'' to learn more about how agencies can use social media as a means of shaping their reputation. Gradient boosted trees have been successfully applied to classification problems in a wide variety of contexts and have the added benefit of being among the most transparent and interpretable machine learning algorithms currently available~\citep{chen2016xgboost,kleinberg2017human}  This application is primarily for demonstration purposes, although this procedure can be applied to an expanded social media database to gain much broader understanding about how agencies use social media to shape their reputations.

\subsection{Measuring Organizational Reputation}

 There are several different ML and non-ML based approaches which could be used to measure organizational reputation, either through using different sources or taking a different approach to the measurement of organizational reputation described here.
While previous approaches to organizational reputation have focused on measuring the extent to which the public and others perceive the reputation of public organizations, challenges related to measurement and volume of organizational data have prevented researchers from studying organizational reputation from the inside out.
That is, studying how public organizations attempt to strategically \textit{shape} their reputations and what the rationales for and consequences of, these strategies entail. 
It is in the area of \textit{shaping} organizational reputation that we believe ML methods hold the most promise for in public administration research.
Below we discuss some of the previous, non-ML approaches in greater detail,  discuss other ML-based approaches that we could have taken to measuring organizational reputation and finally provide an in depth discussion of the supervised approach that we take in this paper. 

In the literature, experimental approaches and proxies from administrative data have been used to measure public perceptions of organizational reputation.  For instance, in a study exploring the effects of organizational resources on agency performance,~\cite{lee2012assessing} measure organizational reputation using publicly available Freedom of Information Act (FOIA) data. Specifically,  they combine the number of FOIA request denials and the time to respond to a request as a proxy for agency reputation, arguing that how the agency is perceived by the public will be shaped by these metrics.  In a recent paper by~\cite{teodoro2018citizen}, the authors apply a model of consumer branding to the public administration context and use a series of experiments to gauge brand favorability among a group of four federal agencies. 

As mentioned above, these approaches both measure organizational reputation as a function of how citizens and the public \textit{perceive} organizations, but little is known about how organizations strategically \textit{shape} their own reputations through media and other materials that they distribute to the public. It is for these purposes that we believe ML holds the most promise by allowing public administration researchers to extract and codify organizational data at a scale that was simply not possible in the past. While we pursued a \textit{supervised} ML approach, there are other ML-based approaches which researchers might consider either using unsupervised ML or taking an alternative supervised approach to the one we took here. 

An unsupervised approach that we had considered here, but decided against, was one in which we would have utilized topic models. Here the approach would have involved estimating a series of topic models using agency tweets and identification of aspects of  categories within topics in each of the models as a means of teasing out~\cite{carpenter2012reputation}'s reputation categories.  One of the most important considerations preventing us from taking this is approach is that topic models are typically ill suited  to shorter texts and   produce uninterpretable topics with these kinds of data~\citep{roberts2014structural}.
An alternative supervised ML approach that we could have taken, and one which future researchers may be interested in taking, is a supervised approach using a framework developed by~\cite{waeraas2012public}.

In this paper~\cite{waeraas2012public} identify five problems facing public organizations in their attempts to engage in reputation management. These include (1) the politics problem (2) the consistency problem (3) the charisma problem, (4) the uniqueness problem, and (5) the excellence problem. Tweeting by agencies can, in this light be seen as a means of potentially reflecting and addressing one or several of these problems. Thus, an alternative supervised ML paper which could have been written around these concepts might involve using supervised ML to identify which whether tweets written through agency accounts address either of these problems.  

In this paper, we took the approach that we did because it is best suited to the task of directly measuring the theoretically broadest types of organizational reputation discussed in~\cite{carpenter2012reputation} from texts, a task which effectively involves mapping an abstract set of categories onto agency tweets. 
In particular, our thought process behind the methodology we employed was based on the premise that the best we could do, measurement wise, for this task was to attempt to accurately reproduce human coding efforts in the face of a lack of specific a priori guidance regarding how this classification task should be accomplished.


Outside of historical analysis, there are limited data for understanding how agencies use social media to shape their reputations. The first task is to construct an original database for the study of how agencies can use social media for communicating reputational information to the public. The steps in this process for linking the text data that we can acquire about agencies through social media (i.e., tweets and posts) to aspects of reputation identified by ~\cite{carpenter2012reputation} are: (a) Conceptualization; (b) Data collection; (c) Human coding; (d) Machine learning algorithm training and testing; and, (e) Analysis.\footnote{ \textit{R} code accompanying each of these steps can be found in the Appendix.}

\subsection{Conceptualization}

 \textbf{Conceptualization} involves identifying textual content relating to organizational reputation using social media data. For our purposes, the unit of analysis for social media data generated by federal agencies is a short piece of text in the form of a tweet (or post) on the Twitter social media platform. Our goal is to identify textual content within each of these posts that relates to organizational reputation. We conceptualize classification as labeling each tweet as containing one of the four types of reputation identified by ~\cite{carpenter2012reputation} (performative, moral, procedural, and technical) or containing content unrelated to any of these. Table~\ref{tab:reputation} contains descriptions of each type of reputation.

\begin{table}[ht!]
\centering
		\begin{tabular}{ll}
	\hline\hline
			\textbf{Reputation Type} &  \textbf{Description} \\ \hline
			Performative & ``Can the agency do the job?''  \\
		    Moral & ``Does the agency protect the interests of its clients,\\
		    			&		 constituency and members?'' \\
		    Procedural & ``Does the agency follow accepted rules and norms..'' \\
		    Technical & ``Does the agency have the capacity and skill required \\
		    				&  for dealing in complex environments...'' \\ \hline\hline
		\end{tabular}
		\caption{Four types of organizational reputation for labeling social media posts}
		\label{tab:reputation}
\end{table}

\subsection{Data Collection}

After establishing a coding scheme, we \textbf{identified the data for collection and the agencies from which to collect data}. This application includes data from 13 executive agencies in the Cabinet of the President of the United States  that have a presence on the Twitter social media platform (see Table~\ref{tab:agencies}). Using the Twitter API (application programming interface), to ensure that we had a balanced sample across agencies and to account for natural differences in the volume of agency tweets,  we collected approximately 2,000 tweets per agency that were available on Twitter as of March 28th, 2018 by connecting using the \textit{R} programming language and extracting the textual information from tweets from each agency. We collected 26,402 tweets. More information about each agency's Twitter feed can be found in Table~\ref{tab:agencies2}.

\begin{table}[ht!]
\centering
	\begin{tabular}{ll}
	\hline\hline
		\textbf{Agencies} &  \\ \hline
	Health and Human Services	 & Education \\
	Housing and Urban Development	 & Interior \\
	Homeland Security	 & Treasury \\
	Defense	 &  Agriculture \\ 
        Transportation      &   State  \\	
	Commerce	 &  Justice\\ 
        Energy        &   \\	
		\hline\hline
	\end{tabular}
	\caption{Included agencies}
	\label{tab:agencies}
\end{table}


\subsection{Human Coding}
\textbf{Human coding of data} is one of the most important stages of building any machine learning model. All supervised machine learning algorithms ultimately rely on the quality of the human--coded data as inputs, so care must be taken in sampling the data that are presented to coders and in the coding instructions. As with endeavors involving traditional statistics, a representative sample requires that samples be randomly selected from a sampling frame. In this application, the sampling frame was the 26,402 tweets from the collection of agencies. From this, 200 tweets were randomly sampled and presented  to three individuals; two were Mechanical Turk ``masters'', coders without expertise in the area,  and the other was one of the authors of the paper, an ``expert'' coder with background knowledge of the issue.  Instructions given to the Mechanical Turk workers were identical to the criteria used by the author. That is, each used only the definitions of organizational reputation provided by Carpenter and Krause (2012) to determine which category each of the randomly sampled 200 tweets fell into.  Instructions provided to the coders are provided in the Appendix. While it is certainly realistic to believe that a tweet can belong to multiple reputation  categories, the constraints imposed on us by the machine learning algorithm require that we only have a single class label (reputation category) that we can use to train the algorithm on.  As a result, we must constrain coders to consider which reputational category is the ``best fit'' for the tweet given the available information. 

Table~\ref{tab:tabhand} contains the distribution of the reputational categories identified by the hand coding exercise as conducted by the expert coder. Accurate predictions using ML algorithms require as much data as possible, so we use the moral reputation category to train our algorithm for this application. Here are examples of tweets that were coded as containing content related to moral reputation, including one from Homeland Security that celebrated new citizens: \textit{``\@ USCIS this morning welcomes \#newuscitizens at the \@USNatArchives in honor of \#ConstitutionDay.''} Another from Veteran's Affairs affirmed the agency's support for veterans: \textit{``Still supporting Veterans, a half million video views later''} . Finally, one from the Department of Education expressed concern about homeless students and what the agency was doing to help them: \textit{``This year, we heard from students experiencing homelessness on how to better support their needs''}. 

\begin{table}[ht!]
\centering
	\begin{tabular}{ll}
	\hline\hline
		\textbf{Category} & \textbf{\% of Tweets}  \\ \hline
		Performative &  12\% \\
		Moral & 32\% \\
		Procedural & 1.5\%  \\
		Technical &  12\% \\ 
        None & 42.5\%	\\	
		\hline\hline
	\end{tabular}
	\caption{Tweets by reputational category}
	\label{tab:tabhand}
\end{table}

The instructions given to the coders were purposefully vague and did not prompt coders to search for ``key words'' which might relate to each of the reputation categories (see below). This was done because first, it could not be entirely clear what words mapped onto each of the reputational categories and second, prior ML research on sentiment analysis has found that allowing individuals to make classification decisions on the basis of more abstract rules tended to result in ML classifiers that did a better job of predicting the categories. 
We recognize that some might dispute this approach. We will discuss two aspects of this approach that are particularly important for considering the value of machine learning. First, just as with traditional coding, the human coding component can be handled in a way to check for reliability and coder agreement. We want to note in this case, though, that the nature of ML (as described in our first draft) is built on the notion that checking for algorithmic agreement is equally important. Second, because ML is relatively fast compared to human coding, it is easy to vary human coding approaches and then rebuild the ML estimates based on alternative human coding approaches. 

\begin{figure}
\caption{Intercoder reliability between non-expert coders and an expert coder}
\label{fig:intercoder}
\centering
\includegraphics[width = 0.8\textwidth]{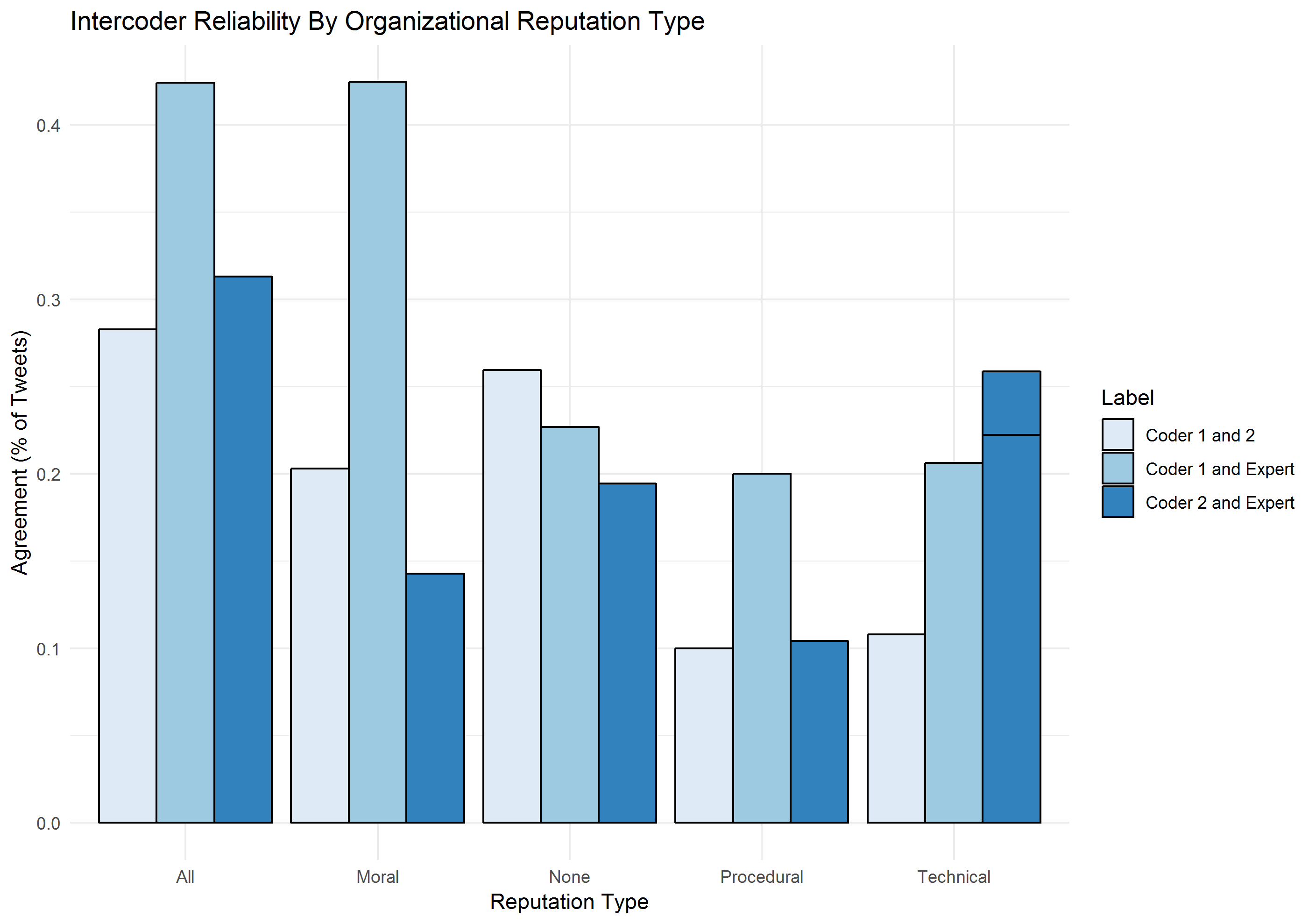}
\end{figure}

To highlight some of the challenges inherent in coding complicated data such as these, we estimate intercoder reliability between each of the non-expert coders and each of the non-expert coders with the expert coders.  Figure~\ref{fig:intercoder} plots intercoder reliability as a function of the proportion of tweets that each of the coders agreed upon overall and intercoder reliability by category.  From this plot, we see that agreement between non-expert coders was poor overall and by category but that agreement between one coder and the expert was significantly higher overall and higher within the category of ``moral reputation''.   For these cases, where intercoder reliability is low, we recommend that algorithm training \textit{always be done using data labeled by expert coders} and that reliance on non-experts be kept at a minimum. This is particularly important during the performance measurement stage of algorithm training as algorithms trained on data coded by experts who are knowledgeable about the subject matter and who take care and time to do the coding, will, with rare exceptions, always be superior to those trained on data labeled by non-experts. As a demonstration of this, we train the algorithm below using labeled data from all three coders, and find that performance is maximized by using the data labeled by the expert coder. 

\subsection{Machine Learning Algorithm Training and Testing}

After acquiring the hand-coded data, the next step is to determine whether a machine learning algorithm can be legitimately employed to code tweets in the larger database. Given text data, this process of \textbf{training and testing our algorithm} requires several extra steps. The general process is as follows: 

$$
\textbf{Text Pre-processing}  \rightarrow \textbf{Training} \rightarrow \textbf{Testing} \rightarrow \textbf{Performance}
$$

\noindent First, we chose the machine learning algorithm(s) for building the classifier. This algorithm is ``gradient boosted trees'', which is known to perform very well on text classification tasks~\citep{si2017gradient}; it is also one of the most transparent and interpretable algorithms, which allows us to learn more about \textit{how} the classifier was able to distinguish between tweets containing content relevant to moral reputation and those that do not. Second, the data were divided by randomly shuffling the data and partitioning it into two datasets: one for training and a second for testing. In this application, 70 percent of the data were reserved for training and 30 percent for testing; this is a common default train/test split in most automated statistical packages and is frequently recommended when data are sparse. 

Because we are dealing with text data, we must \textbf{pre--process the text} by preparing the raw text of the tweets for analysis and converting them into a matrix called a document--term matrix, which has the number of rows equivalent to the number of documents in the training data and the number of columns equivalent to the number of terms, or unique words and phrases, within \textit{all} of the documents in the training data.

Pre--processing the tweets is first performed to maximize performance of the machine learning algorithm~\citep{grimmer2013text,hollibaugh2018use,denny2018text} using a series of natural language processing algorithms embedded in a function which we defined as ``text\_cleaner()'' in the  \textit{R} statistical language. These functions are often referred to as ``pipelines" in the ML community because they use a series of steps to pass ``dirty'' text through a function to produce ``clean'' text. Pipelines often must be tailored for each specific type of text used for analysis. In our case, the pipeline that we passed our text through involved: \textit{tokenization} (or separation) of the words in the tweets into unigrams (single words); \textit{removal of stop words} (common words like ``the" that add little information); \textit{removal of all special characters, numbers, and punctuation}; and, \textit{word stemming}, in which suffixes (``-ing'', ``-s'', ``-ed'', etc) are removed and replaced with the word stem. The pipeline ensures the algorithm can achieve the best possible performance given any set of texts.\footnote{The ``text\_cleaner()'' function is included in the code for this paper.}

After pre--processing, data were automatically converted into a document--term matrix using the ``tm'' package in \textit{R}. The rows of the document term matrix contains the tweets; the columns contain vectors corresponding to each word in the documents. Each element of this matrix contains a measure of the text frequency -- counts of each of the words in the tweet.\footnote{Alternatives methods for weighting document term matricies such as TF--IDF (or text frequency/inverse document frequency) are beyond the scope of this paper, but are discussed in more detail in~\cite{hollibaugh2018use}.} Each tweet contains far fewer words than all of the unique words in the corpus so each row of the document--term matrix is mostly zeroes; it is a \textit{sparse} matrix. 

The next steps are \textbf{training the algorithm, testing it, and assessing its performance}. In the parlance of the ML basics reviewed above, the document--term matrix was the set of \textit{features} used to predict the \textit{class label}; in this application, the class label is a binary variable indicating whether a tweet contains content related to moral reputation. The intuition behind the gradient boosted tree classifier is relatively straightforward, although the details may not be. Imagine that for each word in a document, one wanted to create a hierarchy of which words were most and least important in helping distinguish tweets which contain moral reputation content (versus those that do not). To construct this hierarchy, one must first figure out, for each word, what proportion of tweets relating to moral reputation contain each word. 

If 60 percent of ``moral reputation'' tweets contain the word ``learn'' (while 40 percent do not), this implies that ``learn'' is very good (in itself) for identifying tweets related to moral reputation. ML often uses measures called ``entropy'' and ``information gain'' from information theory to calculate how ``good'' each word is (on its own) for identifying the class a tweet belongs for each word. Once the word with the highest information gain is computed (e.g., ``learn''), that word takes the highest position in the tree. A ``split'' is then made to determine (conditional on containing or not containing the word ``learn'') the other words that help further distinguish between tweets of each category. The process of splitting and information gain calculation is performed iteratively until all of the words in the collection of tweets are structured as a hierarchical tree in which the most important words relevant to predicting moral reputation are at the top and the least relevant are at the bottom.

\begin{figure}[ht!]
\centering
\includegraphics[scale=0.4]{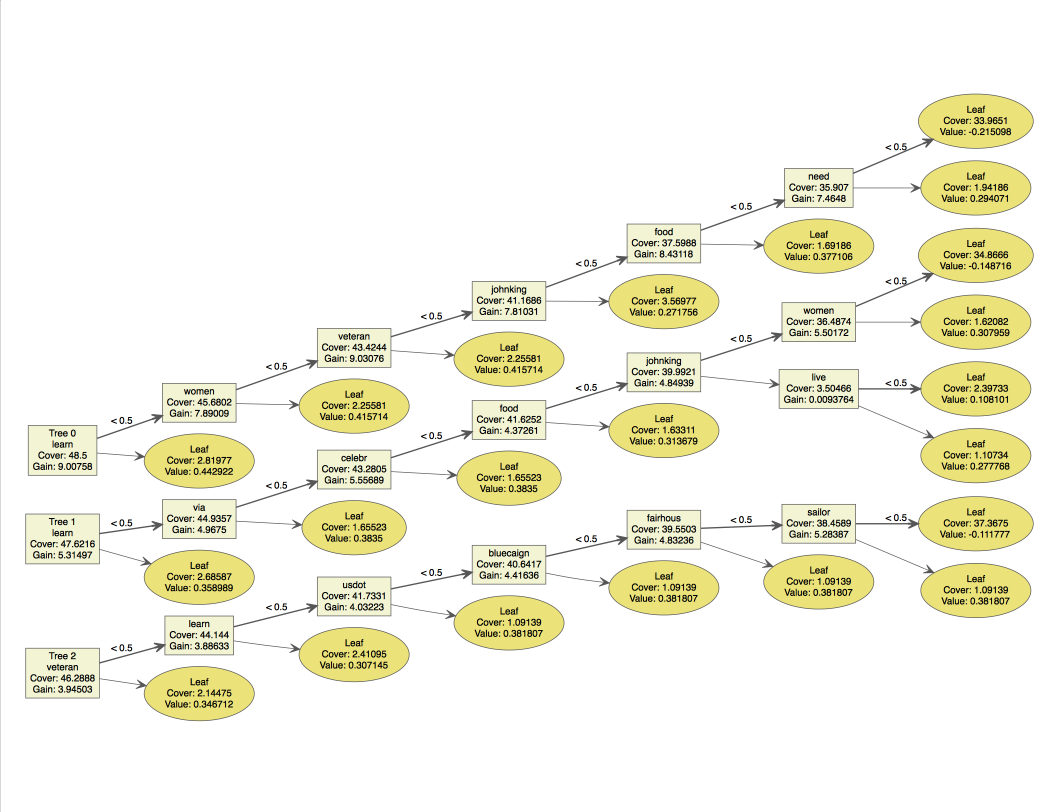}
\caption{Trees produced by the trained gradient boosted tree algorithm}
\label{fig:tree}
\end{figure}

Figure~\ref{fig:tree} contains an example of a portion of three trees in which the words ``learn'' and ``veteran'' are clearly the most important in determining moral reputation in tweets. There is more than one tree in this application because the gradient boosted tree algorithm estimates multiple trees using random samples of the training data to build the classifier. Iteratively sampling the data (and building trees based on these samples) avoids \textit{overfitting}, which occurs when a full tree containing all of the terms is grown. After several trees are grown, when the classifier is introduced to new data, each tree will predict what the class label should be, but the final judgment about that class label depends on a ``majority vote'' of the trees. Specifically, the trees in Figure~\ref{fig:tree} were constructed through an iterative process in which each of the features/independent variables were split into two groups and the gain in information (the measure of the ability of the classifier to predict the outcome/class label) was calculated for each of these splits. This figure shows that the words that produced the most information gain were ``women'', ``via'', and ``learn'' across three different trees estimated.

Suppose we trained a classifier that grew seven trees and presented the text of the following tweet to the classifier: ``This year, we heard from students experiencing homelessness on how to better support their needs.'' Each tree would produce a predicted probability that the tweet contains content related to moral reputation given the words, or $p(Moral | Words)$. If $p(Moral | Words) > 0.5$, then the tweet is classified by each tree as being a ``moral reputation'' tweet. If the majority of the seven trees (perhaps trees numbered $1/7$, $2/7$, $3/7$, and$4/7$) classified the tweet as a ``moral reputation'' tweet, then the final label is ``moral reputation'' (the converse is also true). Table~\ref{tab:majority} provides an example of a majority ``vote''. 

\begin{table}[h!tb]
\centering
\begin{tabular}{llll}
	\hline\hline
\textbf{Tree} & \textbf{$p(Moral|Words)$} & \textbf{Label} & \textbf{ Final Class} \\ \hline
1   		& 0.60    & 1 & Moral reputation \\ 
2			&0.55    & 1& (57\% majority)\\ 
3			& 0.80   &  1\\ 
4			& 0.70   & 1\\ 
5			& 0.45    & 0 \\ 
6			& 0.32    & 0 \\ 
7			& 0.39    & 0 \\  \hline\hline
\end{tabular}
\caption{Example of classification using gradient boosted trees}
\label{tab:majority}
\end{table}

Once a gradient boosted tree model learns how to distinguish between tweets using the training data, the model is applied to the test data to make predictions about tweets that the model has not seen, thereby simulating the process of applying the model to data that are ``in the wild''. Comparing the model's predictions with the human--coded labels reveals if the trained classifier can distinguish between tweet types as well as a human coder. In this application, the model was trained by randomly dividing the data. After the algorithm trained on 70 percent of the data, the remaining 30 percent the data was then passed through the trained model to generate predictions. Quality of the machine generated predictions was assessed by comparing the predictions against the human--coded data as a baseline. 

To illustrate the importance of using data generated by expert coders,  we trained three models produced by each of the three coders: the two, non-expert Mechanical Turk workers and the expert coders and then assessed model performance based on each of these individuals. 
As we demonstrate below, the model trained on data produced by the expert coder performed significantly better than the model trained on data produced by either of the non-experts. In most situations when dealing with complex data that require expert knowledge this is to be expected. Since all supervised machine learning algorithms are ultimately designed to replicate human coding as best as possible, when care and subject matter knowledge are used by experts to code data, this will be almost always\footnote{Here was say ``almost always'' because there is no mathematical theorem or result that can guarantee this outcome, but when greater care is taken by a coder to perform the coding task, the increases the likelihood that the algorithm will be able to pick up on textual information the more clearly distinguished the categories of interest.} be reflected in the quality of the trained algorithm as measured by performance metrics.

Tables~\ref{tab:confusion} and~\ref{tab:perform} both contain means of measuring the performance of the algorithm. Table~\ref{tab:confusion} summarizes the results of applying the trained model on the test data for the ; the \textit{confusion matrix} records which tweets the classifier and the expert human coder disagreed on most (the counts of correctly identified tweets by the classifier are on the diagonal). In this case, the classifier is better at predicting the ``Other'' category than the ``Moral Reputation'' category. Yet, the confusion matrix \textit{per se} does not reveal much about the classifier's overall performance in terms of distinguishing between tweets. The counts in the confusion matrix help measure accuracy, precision and recall -- through the $F_1$ score ~\citep{yang1999evaluation}.

\begin{table}[htbp]
\centering
\begin{tabular}{r|rr}
	\hline\hline

  			 & \multicolumn{2}{c}{\textit{Expert Coding}} \\ \hline
	\textit{Classifier}		 & 	\textbf{Other}	&  \textbf{Moral} \\
  \hline
\textbf{Other} 						& 	29 &   16 \\ 
  \textbf{Moral} & 2 & 13 \\ \hline\hline
\end{tabular}
\caption{Expert coder confusion matrix for classification of tweets}
\label{tab:confusion}
\end{table}

Table~\ref{tab:perform} presents these metrics for the algorithm trained on data coded by an expert and two non-experts .  Accuracy is the percentage of correct predictions made by the classifier on the test data. The ``no information rate'' is the probability of successfully guessing the correct categories by chance. Precision gives us a sense of how good an algorithm is at distinguishing between true positives and false positives. It is calculated as the fraction of true positive results divided by the number of true positives plus false positives. In this case, true positives are tweets identified by the human and the algorithm as being related to moral reputation and false positives are tweets identified by the algorithm as being related to moral reputation which, according to the human, are not related to moral reputation. 

It is easy to calculate the precision of the algorithm trained on the expert labeled data using the confusion matrix from~\ref{tab:confusion}. The total number of moral reputation tweets that the expert and the algorithm agreed on are 13 and the total number of false positives are 2. Thus, the \textit{precision} of the algorithm using the expert data is 13/15 = 86 percent. \textit{Recall} on the other hand, gives us a sense of how good an algorithm is at detecting cases labeled by the human as positive, in this case, moral reputation tweets as labeled by the expert coder. It is simply the ratio of the number of  moral reputation tweets that the coder and the algorithm agreed on (13) divided by the total number of moral reputation tweets as identified by the human coder (29). Thus, the \textit{recall} of the algorithm trained on the expert coded data is 13/29 = 41 percent. To understand both of these numbers conceptually, these statistics are telling us that when this algorithm identifies a moral reputation tweet, we are roughly 86 percent certain that this tweet is truly related to moral reputation (precision). However, the algorithm will tend to \textit{underestimate} the number of moral reputation tweets in any given Twitter database, identifying approximately only 41 percent of moral reputation tweets (recall). Finally, the $F_1$ score combines precision and recall to give a more complete sense of the algorithm's overall performance for both of these tasks and is defined mathematically as normalized ratio of precision and recall\footnote{The $F_{1}$ statistic is simply: $2 \times \frac{precision*recall}{precision + recall}$ }.

\begin{table}[htbp!]
\centering
\begin{tabular}{llllll}
	\hline\hline
				 &	\textbf{Accuracy} & \textbf{No Info. Rate} &	\textbf{Precision}	&  \textbf{Recall} & \textbf{$F_1$}  (0-100) \\ \hline
\textbf{Expert} &  69.1\% & 51.7\% & 86.7\% & 44.8\% & 59 \\ 
\textbf{Non-Expert 1} & 46.7\% & 58.3\% &  28.6\% & 8.0\% & 12.5 \\
\textbf{Non-Expert 2} & 47.1\% & 86.7\% & 0.0\% & 0.0\% & 0 \\
\hline\hline
\end{tabular}
\caption{Performance statistics for algorithms trained with different coders}
\label{tab:perform}
\end{table}

Table~\ref{tab:perform} contains performance statistics described above for the algorithm trained on data coded by the expert and each of the non-experts.  Here we can clearly see through these statistics that the expert coded data is of higher quality than that of the non-experts. The stark differences in these statistics across the board demonstrates the importance of careful coding by individuals with substantive knowledge of complex issues such as organizational reputation. Focusing on the most relevant and important statistics for measuring algorithm performance, precision and recall, we observe that the algorithm trained using the expert data has more than double the precision and more than triple the recall of the algorithms trained on data generated by non-expert Mechanical Turk ``masters.''

\subsection{Analysis}

Given a ``trained'' classifier that performs well at distinguishing between tweets with moral reputation content, the classifier was then applied for the \textbf{analysis} of the larger database of tweets to learn how agencies use social media in the context of shaping organizational reputations. The first goal was to create a prediction machine from a relatively small amount of data (200 observations) for the purpose of learning about content in a much larger database of tweets collected across agencies. To do so, the trained model was applied to the remaining 26,202 tweets to predict whether each tweet contained moral reputation content $M^{*}$. This prediction was based on the estimated probability that each tweet contains moral reputation content, or $p(M_{i} | W^{(i)})$. This probability was estimated from the tweet text via the gradient boosted tree model. Formally, the text of each tweet $i$ was passed through the classifier and labeled using this rule: for each tweet $i$, label the tweet as containing moral reputation content when $p(M_{i} | W^{(i)}) > 0.5$. Using this rule, we labeled each of the 26,202 tweets and computed the mean percentage of moral reputation tweets for each of the 13 agencies.

\begin{figure}[ht!]
	\centering
	\includegraphics[width = .9\textwidth]{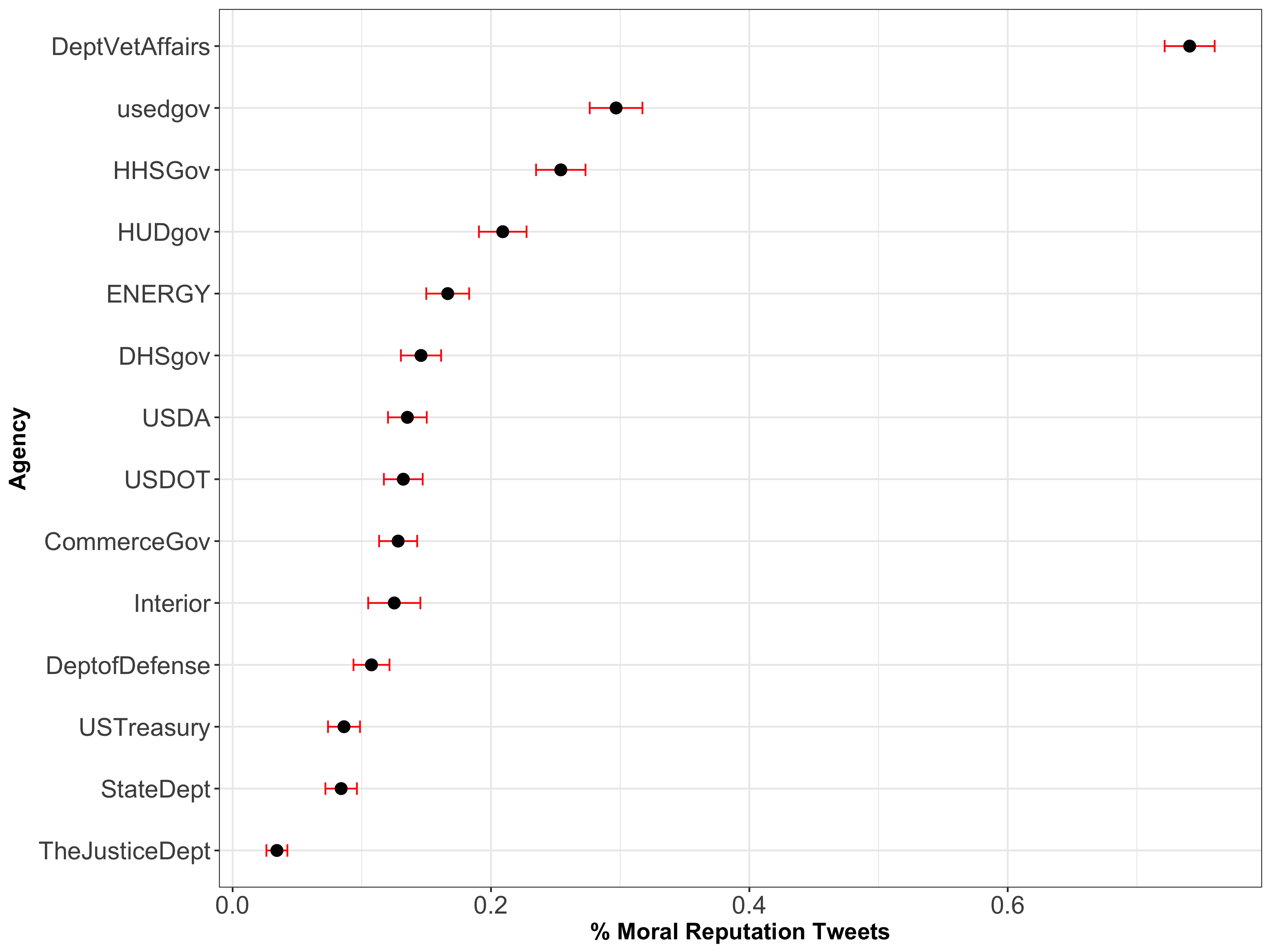}
	\caption{Mean estimated percentage of moral reputation tweets by agency}
	\label{fig:moralrep}
\end{figure}

Figure~\ref{fig:moralrep}  shows this percentage by agency. Only 200 tweets were hand--coded; moral reputation was predicted in the remaining 26,202 tweets. Specifically, the dot is the mean estimated percentage of moral reputation tweets for each agency; the plot shows the ranks from highest to lowest using the full 26,402 tweet database collected, along with 95 percent confidence intervals. 

The Department of Veterans Affairs and the Department of Education stand out as containing the highest percentage of tweets expressing moral reputation. We might conclude that agencies charged with directly serving the general public (Education, HHS) or targeted constituencies (Veterans Affairs) appear concerned about their moral reputation and utilize social media as a means of projecting moral reputation to the public. On the opposite end of this spectrum, agencies that may rely more heavily on internal political support (such as the DOJ, Interior, and the State Department) appear to have little expression connected to their moral reputation. This is intriguing regarding which agencies are concerned with moral reputation, but also lends some face validity to this supervised machine learning classification exercise. 

\begin{figure}[ht!]
	\centering
	\includegraphics[width = .9\textwidth]{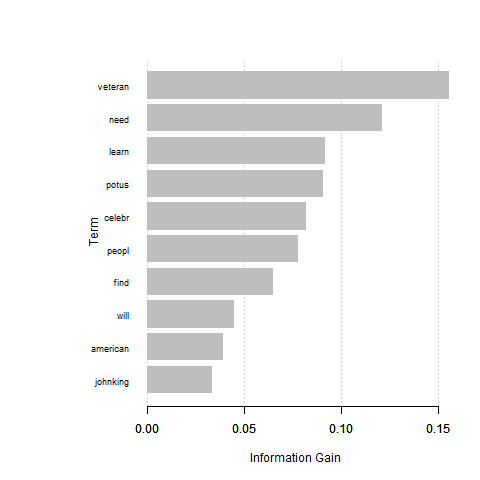}
	\caption{Top words for distinguishing tweets}
	\label{fig:termimport}
\end{figure}

The gradient boosted tree algorithm also automatically provides a means of assessing the words that were most important in distinguishing tweets, which offers insights into the linguistic features of tweets related to moral reputation. Figure~\ref{fig:termimport} plots the top ten most important terms used to distinguish between tweets (based on average information gain across trees). We can see that ``veteran,'' ``need'' and ``learn'' were most relevant in distinguishing between tweets containing moral reputation content.  This is to be expected since moral reputation tweets were tied most strongly to those coming from veterans affairs and education.

\section{Perils and Promise of ML Algorithms}

What are the prospects for machine learning in the public administration context? Should we see thinking machines (in Simon's terms) as a way of handling the deluge of data that threatens to overload human attention? 

The purpose of this section is to offer three perils and three promises associated with the advent of ML algorithms in the linked worlds of public administration research and practice. Before offering these six comments, we summarize the inertia behind the ML enterprise by offering an example. Imagine a human coder had coded the remaining 26,202 tweets. How long would it have required? If that coder required one minute (on average, given inefficiency and human limits) to code a single tweet, the remaining 26,202 would have required 437 hours -- or 55 eight-hour workdays. (Given a research assistant limited to 13 hours per week, that would be 34 weeks -- over two semesters worth of coding.) The point of this simple demonstration is to show the substantial demand for ML algorithms.

\begin{center}
	\textit{Peril 1: Data are everything.}
\end{center}

The ability of machine learning algorithms to generate predictions depends entirely on the data that they use to make these predictions. Because of this, if predictions are made using data that are are of poor quality or are not representative of a population, predictions will also be of poor quality and unrepresentative of a population. The old saying ``garbage in, garbage out'' is particularly relevant in the context of machine learning algorithms.

In this context, it is important to reiterate that because of their speed, ML algorithms can be estimated many times using different input data. The example we offer here shows the importance of using expert-coded data. Even Mechanical Turk masters are poor coders! One might argue that even academics are poor coders for certain topics. If so, then truly expert coders (e.g., practitoners) might be employed for the small sample that we use as the input for the ML algorithm - an opportunity unlikely to present itself if we wanted an expert to code 26,000 tweets. 

\begin{center}
	\textit{Peril 2: Correlation does not equal causation.}
\end{center}

Machine learning algorithms are optimized to make good predictions, but this fact alone is simply not sufficient to justify policy decisions, or other types of actions, based solely on their predictions. At the end of the day, these algorithms are fundamentally tools that only guarantee very good correlations.

\newpage 

\begin{center}
	\textit{Peril 3: Good predictions often require a tradeoff between accuracy and interpretability. }
\end{center}

As scholars of human behavior and organizations, we are often more interested in \textit{why} we are able to make predictions by understanding which variables in our data were most relevant in terms of our ability to generate accurate predictions.  While some algorithms -- such as gradient boosted trees, random forests, naive Bayes and regularized logistic regression -- are highly interpretable ``white box'' methods, other methods such as neural networks are ``black boxes'' that are practically impossible to interpret. 

Indeed, one of the most significant challenges faced in this domain is a tradeoff that must be made between predictive accuracy and interpretability as models that tend to perform better in terms of predictive accuracy, such as neural networks, often tend to be the least interpretable. As scholars of individual behavior and of organizations, we should \textit{always} first consider the most interpretable models before applying ``black box'' methods to a difficult prediction problem that we wish to solve. Despite these pitfalls, however, machine learning algorithms also hold a great deal of promise both for organizational decision making and as research tools.

\begin{center}
	\textit{Promise 1: Variety and veracity of big data.}
\end{center}

Machine learning algorithms are able to address two of the three Vs of big data: variety and veracity. They are able to handle a wide variety of data forms including texts and images in an efficient manner. Through  dimensionality reduction methods such as ``regularization,'' machine learning algorithms provide a means by which researchers can sort through millions of variables to determine which are important.

\begin{center}
	\textit{Promise 2: Ethical model selection}
\end{center}

One of the biggest ethical challenges facing modern social science research today is the practice of ``p--hacking,'' in which researchers consciously manipulate the statistical models that they estimate and the data that they choose to include in order to ensure that they achieve a statistically significant ($p < 0.05$) result under a null hypothesis of no effect. In theory, machine learning algorithms allow researchers to conduct principled model selection in situations in which the risk of p--hacking is greatest: when faced with hundreds or even thousands of variables to choose from and where little theoretical guidance exists. 

We also note that there ethical model selection might also include broader concerns about the identity of the expert coder. In some ways, these coding decisions are part of the ``super-model'' -- the broader aspects of how we build and assess statistical models -- that we often gloss over when we debate technical concerns like Xs and Ys and the order of their inclusion in the model.

\begin{center}
	\textit{Promise 3: Data driven decisions for public organizations and public administrators.}
\end{center}

Machine learning algorithms provide public administrators with a resource that can help them make important administrative decisions in a consistent and rapid manner. For example, immigration officials at the U.S. Citizenship and Immigration Services (USCIS) across the United States are responsible for making decisions about a variety of visa applications every day. In theory, machine learning algorithms could assist these individuals by providing them with a score which utilizes all of the information about the visa applicant in the application to determine the likelihood that an application will be denied or accepted. Officials can then use this information to focus more attention on scrutinizing applications that are likely to be denied and less attention on scrutinizing applications that are likely to be certified~\citep{kashyap2017not}. 

\section{Conclusion}

Machine learning algorithms are the building blocks for prediction machines. These machines make it possible for a computer to learn about the essential features of abstract organizational concepts such as reputation and communication from texts. Moreover, researchers can use prediction machines to reproduce codings without the need of human coders. 

While this paper has focused on the process and methods of machine learning, we want to close this discussion by focusing on a core aspect of ML that is inherent to the approach: it is automated. In the introduction of this paper, we emphasized Simon's concern about data swamping human attention -- that humans process information serially so we cannot keep up with its ever-increasing flow. For Simon, thinking machines might help with this imbalance. One contribution of this paper is to help review some of the technical details for this conceptual problem.  Of course, as we discuss here, all technical solutions carry both promises and perils. In the long run, though, the amount and velocity of data mean the inevitable automation of statistical analysis; there is a ``machine'' in machine learning.    

Where does this end? Already practitioners have looked beyond machine learning applications like the one we presented because in many ways the process requires too much human intervention. ML is a family of methodologies that enable but do not fully satisfy the demand for tools that can parse large amounts of complex data. The ML suite requires too much human intervention for a true thinking machine. What is needed is an artificial intelligence that can automate the human parts of ML.

In fact, this is what modern AI relies on: deep learning methods are ML tools on overdrive. Deep learning, expressed as an implementation of ML inside an AI, will revolutionize how governments collect, process, and interpret data. Over time, it may change how public administration researchers engage data as well. Until that happens, this paper's contributions are to provide insight into ML as a research enterprise, to review the basics of ML as a way of ``opening the hood'' on the automated processing of data, and to caution about its use (even as we recognize its promise).

\bibliographystyle{apsr_fs} 
\bibliography{references.bib}

\section*{Appendix}

\subsection*{Instructions provided to coders}

Below are the exact instructions provided to non-expert coders on Mechanical Turk:

{\small \it
For this task, you will view a collection of 200 tweets taken from the twitter accounts of a group of federal agencies and make decisions about how each of these tweets reflects reputational concerns as defined below.

To the best of your ability, please classify each of the tweets according to 1 of the 4 types of organizational reputation categories discussed below. The definitions below are also provided next to each reputation category during the classification task. 

1. \textbf{Performative reputation}: Can the agency do the job? Can it execute charges on its responsibility in a manner that is interpreted as competent and perhaps eﬃcient?

2. \textbf{Moral reputation}: Is the agency compassionate, ﬂexible, and honest? Does it protect the interests of its clients, constituencies, and members?

3.\textbf{ Procedural reputation}: Does the agency follow normally accepted rules and norms, however good or bad its decisions?

4. \textbf{Technical reputation}: Does the agency have the capacity and skill required for dealing in complex environments, independent of and separate from its actual performance?

5. \textbf{None of the above}: The tweet fits into neither category
}

\subsection*{A guide to training the gradient boosted tree supervised learning algorithm}

Included as part of this article is the replication code needed to train the gradient boosted tree model using the \textit{xgboost} package. Since we are dealing with text data and supervised learning, we must employ a total of \textit{three} \textbf{R} packages: 1) the \textit{caret} package contains the 	``confusionMatrix'' function which allows us to calculate performance statistics; 2) the \textit{tm} package contains the tools necessary to clean and transform the text data into a document-term matrix and finally; 3) the \textit{xgboost} package allows us to train and fine-tune the gradient boosted tree algorithm.   Here we outline the steps taken to go from tweet data to a trained gradient boosted tree algorithm the details of which can be found in the code.

\subsubsection*{Step 1: Data cleaning and sparsity reduction}

Using the \textit{tm} package and a function that we wrote for these purposes entitled ``textcleaner.R'', we cleaned the text, passed the text into a document term matrix using the \textit{DocumentTermMatrix} function and reduced the sparsity of the resulting document-term matrix using the \textit{removeSparseTerms} function. Since GBT models are often sensitive to irrelevant information, it is always useful to consider using this function prior to training a model. One caveat of this, however, is that reducing the sparsity too much can result in significant drops in performance, so it is a best practice to not reduce the sparsity of the document term matrix below $95\%$ at least initially.

\subsubsection*{Step 2: Dividing the data into training and test sets}

After preparing the document term matrix, the data was then divided into training and test sets using a 70/30 train/test split. To prepare the data for analysis using the \textit{xgboost} package, additional formatting had to be done using the \textit{xgb.DMatrix} function.  Because of imbalances in the positive and negative class labels, the estimated \% of positive and negative class labels were used as weights in the analyses (\textit{pospredweight} in the code). 

\subsubsection*{Step 3: Training the gradient boosted tree algorithm}

Training the GBT algorithm required two initial steps to fine-tune the algorithm and maximize performance. First a set of parameter values for the algorithm was chosen (\textit{params} in the code) which requires the user to select a boosting method and an objective function. All of the other parameters included were the default values. The objective function chosen was ``\textit{binary:logistic}'' because the outcome variable was binary.  

After parameter values are defined, 5-fold cross-validation is conducted on the training data using the $xgb.cv$ function. The purpose of cross-validation here is to determine the optimal number of iterations (training rounds used to estimate the model) which minimizes prediction error on the test data. After the optimal number of iterations is determined through the cross-validation process, this number of iterations is used to train the final model ($best.iter$ in the code). 

Finally, the model is then trained using the $xgb.train$ function which includes information regarding the parameter values set, the training data, the number of iterations to be estimated and any weights to be used if any. 

\subsubsection*{Step 4: Assessing performance}

After the model has been trained, it was saved as an object in R ($xgb1$). This trained model was then applied to the features in the test data set, predictions were generated and finally the \textit{confusionMatrix} function was used to calculate the relevant performance statistics through comparisons of the predictions made by the model, \textit{xgbpred} and the human coder \textit{testY}.

\subsection*{Agency Tweet Descriptives}

\begin{sidewaystable}[ht!]
\centering 
\resizebox{\textwidth}{!}{\begin{tabular}{lrrl}
  \hline
  Name & Friends & Followers & Description \\ 
  \hline
Agriculture & 2675 & 597708 & Leadership on \#food, \#agriculture.. \\ 
Commerce & 277 & 340324 & Welcome to the official twitter feed of the United States Department of Commerce! \\ 
Defense & 987 & 5645367 & The official account of the U.S. Department of Defense. Following, RTs and links ??  endorsement. \#KnowYourMil \\ 
Education & 151 & 1299122 & News and information from the U.S. Department of Education. \\ 
Energy & 577 & 769871 & Building the new energy economy. Reducing nuclear dangers \& environmental risks. \\ 
HHS & 254 & 729468 & News and information from the U.S. Department of Health \& Human Services (HHS). \\ 
Homeland Security & 383 & 1574626 & The \#DHS Mission: "With honor and integrity, we will safeguard the American people, our homeland, and our values." \\ 
HUD & 358 & 297401 & Creating strong, sustainable, inclusive communities and quality affordable homes for all. Follow does not = endorsement \\ 
Interior & 140233 & 5199258 & Protecting America???s Great Outdoors and Powering Our Future \\ 
Justice & 175 & 1546571 & Official DOJ Twitter account.  \\ 
State & 439 & 4994796 & Welcome to the official U.S. Department of State Twitter account. \\ 
Transportation & 1217 & 190297 & The official Twitter account of the U.S. Department of Transportation. \\ 
Treasury & 257 & 773414 & Executive agency responsible for promoting economic prosperity \& ensuring financial security of the U.S. \\ 
Veterans Affairs & 887 & 629518 & Official Twitter feed of the U.S. Department of Veterans Affairs. \\ 
     
   \hline
\end{tabular}}
\caption{Agency Twitter information}
	\label{tab:agencies2}
\end{sidewaystable}

\end{document}